\documentclass[sigconf]{acmart}

\AtBeginDocument{%
  \providecommand\BibTeX{{%
    \normalfont B\kern-0.5em{\scshape i\kern-0.25em b}\kern-0.8em\TeX}}}

\acmBooktitle{}
\acmPrice{}
\acmISBN{}

\begin{document}

\title{Perceptually Optimizing Deep Image Compression}

\author{Li-Heng Chen}
\affiliation{%
  \institution{The University of Texas at Austin}
}
\email{lhchen@utexas.edu}

\author{Christos G. Bampis}
\affiliation{%
  \institution{Netflix Inc.}
  }
\email{christosb@netflix.com }

\author{Zhi Li}
\affiliation{%
  \institution{Netflix Inc.} }
\email{zli@netflix.com}

\author{Andrey Norkin}
\affiliation{%
  \institution{Netflix Inc.} }
\email{anorkin@netflix.com}

\author{Alan C. Bovik}
\affiliation{%
  \institution{The University of Texas at Austin}
}
\email{bovik@ece.utexas.edu}

\renewcommand{\shortauthors}{Chen, et al.}

\def\x{{\mathbf x}}
\def\L{{\cal L}}

\newcommand{\TBD}{\textbf{TBD}} 
\newcommand{\VMAFp}{$\text{VMAF}\mathrm{_p}$}
\newcommand{\SSIMp}{$\text{SSIM}\mathrm{_p}$}
\newcommand{\MSIMp}{$\text{MSIM}\mathrm{_p}$}
\newcommand{\VIFp}{$\text{VIF}\mathrm{_p}$}
\newcommand\norm[1]{\left\lVert#1\right\rVert}
\newcommand\abso[1]{\left\lvert#1\right\rvert}

\begin{abstract}
Mean squared error (MSE) and $\ell_p$ norms have largely dominated the measurement of loss in neural networks due to their simplicity and analytical properties. However, when used to assess visual information loss, these simple norms are not highly consistent with human perception. Here, we propose a different proxy approach to optimize image analysis networks against quantitative perceptual models. Specifically, we construct a proxy network, which mimics the perceptual model while serving as a loss layer of the network. We experimentally demonstrate how this optimization framework can be applied to train an end-to-end optimized image compression network. By building on top of a modern deep image compression models, we are able to demonstrate an averaged bitrate reduction of $28.7\%$ over MSE optimization, given a specified perceptual quality (VMAF) level. 
\end{abstract}





\maketitle

\section{Introduction}
Deep neural networks have made rapid advances on diverse multimedia tasks \cite{Liu2018, Bosse2018, spaul2019}, especially the image transformation problems including denoising \cite{BurgerSH12}, super-resolution \cite{Lai2019}, frame interpolation \cite{liu2019cyclicgen}, and so on. Specifically speaking, a generative network is learned to reconstruct high-quality output images from degraded input image under a supervised manner. A loss function is defined to measure the fidelity between the output and a ground-truth image. For example, the denoising task aims to reconstruct a noise-free image from
a noisy image, and Convolutional Neural Networks (CNNs) have been shown to provide good noisy-to-pristine mapping functions. However, despite the tremendous amount of research being applied on deep learning image transformation problems, the loss functions used to guide model training has been underexamined and largely limited to the $\ell_p$ norm family. The structural similarity quality index (SSIM) \cite{WangBSS04} has also been adopted as loss functions for several image reconstruction tasks \cite{Snell2017,Zhao2017}, owing to their perceptual relevance and good analytic properties, such as differentiability.

\begin{figure}[!t]
  \centering
  \includegraphics[width=3.35in]{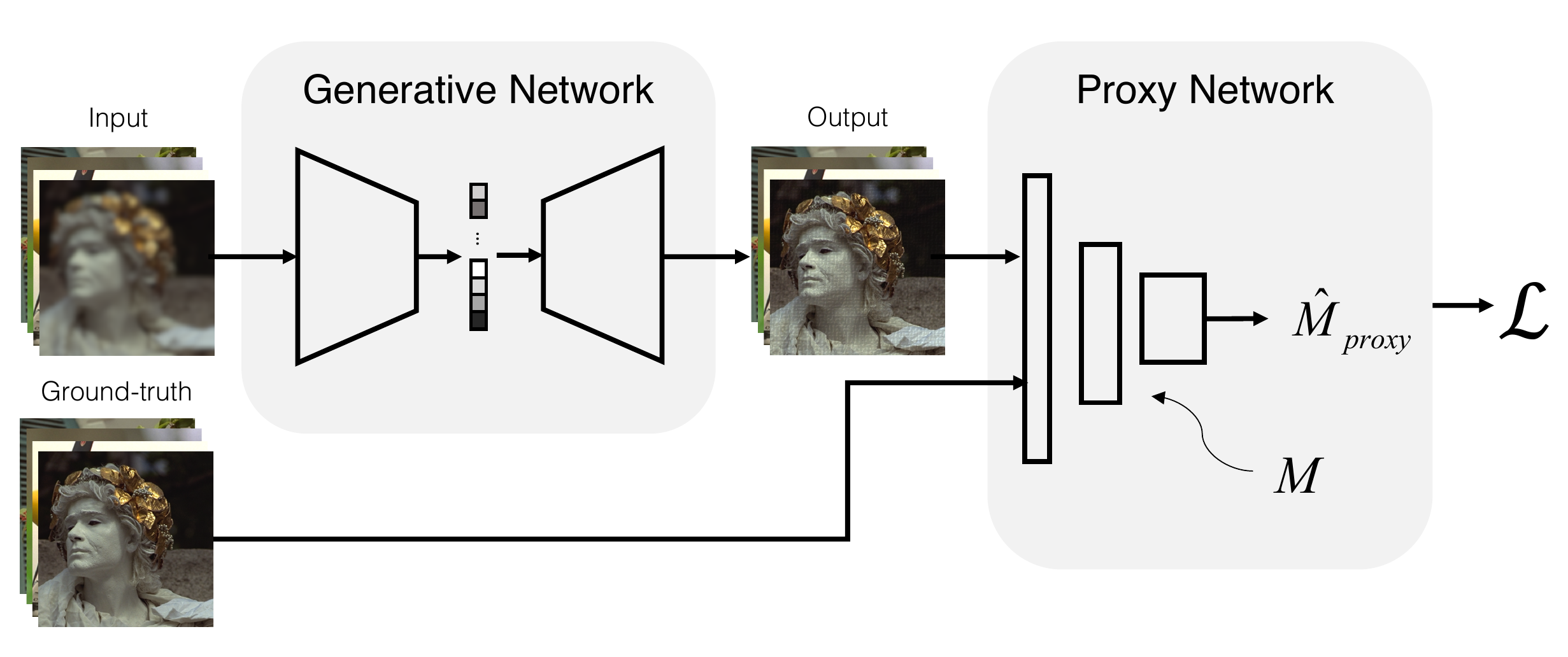}
  \caption{Conceptual framework of perceptual optimization using a proxy network: A generative network takes a mini batch as input, and outputs a reconstructed batch during training. The proxy network is learned as a proxy of an image quality model $M$, where the output $\hat{M}_\mathrm{proxy}$ estimates the quality score predicted by $M$. The generative network learns to maximize $\hat{M}_\mathrm{proxy}$.}
  \label{fig_prox_framework}
\end{figure}

As a long-standing research problem, predicting picture quality with high-quality reference pictures has achieved remarkable success. Numerous full-reference (FR) perceptual models have been proposed and proven to surpass MSE-based measurements. Examples include other SSIM-type methods \cite{WangMSSSIM03, ZhouWangIWSSIM2011,PeiDOG15}, VIF \cite{SheikhB06}, VSNR \cite{DMCSSHVSNR07}, MAD \cite{Chandler2010}, FSIM \cite{LZhangFSIM2011}, and VSI \cite{ZhangSLVSI14}. Moreover, learning based quality predictors such as Video Multimethod Assessment Fusion (VMAF) \cite{ZliVMAF18}, a successful open-sourced example developed by Netflix, has been powerful tools to optimize tremendous volumes of internet video traffic. Unfortunately, most of the advanced, high-performance image quality indeces have never been adopted as loss functions for end-to-end optimization networks, because they are generally non-differentiable and functionally complex.

Recently, lossy image compression models have been realized using deep neural network architectures. This may be viewed as a special case of generative networks, where the input image is equal to the ground-truth image. Unlike the conventional image codecs standards, which rely on ``handcrafted" functional blocks such as transform matrix or in-loop filters, the parameters of learned image compression are optimized in an end-to-end manner. Most of these have employed deep auto-encoders. For example, Ball\'e \textit{et al.} \cite{BalleLS16a} proposed a general infrastructure for optimizing image compression where bitrate is estimated and considered during training. In \cite{balle2018variational}, this model is improved by incorporating a network, scale hyperprior, into the compression framework. The authors use the additional network to estimate the standard deviation of the quantized coefficients to further improve coding efficiency. Later, Minnen \textit{et al.} \cite{NIPS2018_8275} exploit a PixelCNN layer, which they combine with an autoregressive hyperprior. Beyond these early efforts, other recent approaches have adopted more complex network architectures such as recurrent neural networks (RNNs) \cite{Johnston_2018_CVPR}, convolutional autoencoder (CAE) \cite{Cheng2019}, and generative adversarial networks (GANs) \cite{agustsson2018generative}. Some works has also been done to extend these ideas to the deep video compression problem \cite{cheng19}.

In fact, the idea of optimizing conventional codecs such as JPEG or H.264/AVC against perceptual models like SSIM, VIF, or VMAF have been deeply studied \cite{channappayya08,YHH10,WangRWMG12,kslussim20} and implemented in widespread practice \cite{ZliVMAF18}. We seek to both extend this concept, as well as try to bridge the gap between modern perceptual quality models and deep generative networks, we explore the potential of adapting sophisticated perceptual picture quality models as loss functions in deep image compression network. In order to address the aforementioned shortcomings, we conceptually propose to simulate the measurements made by a perceptual image quality model using a proxy network. Then, the proxy network is adopted as a perceptual loss function as interpreted in Fig. \ref{fig_prox_framework}.

\begin{figure}[tb]
  \begin{minipage}[b]{.49\linewidth}
    \centering
    \centerline{\includegraphics[width=4.2cm]{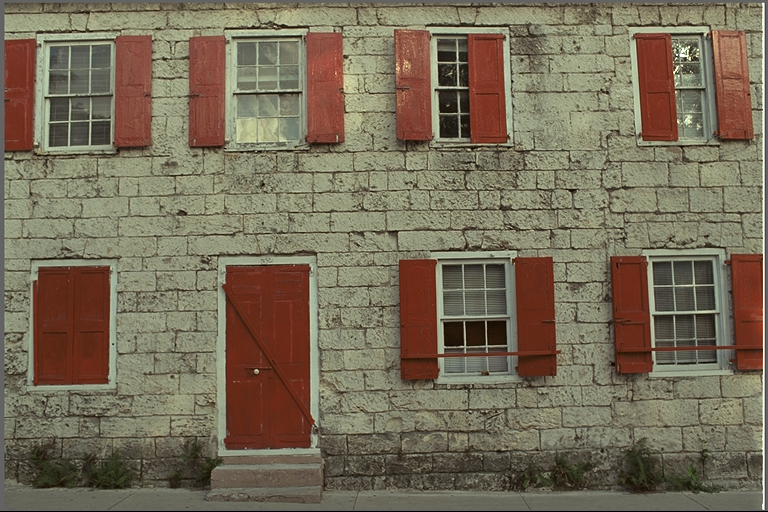}}
    \label{fig_adv_ex_src}
    \centerline{(a) Source image}\medskip
  \end{minipage}
  \hfill
  \begin{minipage}[b]{0.49\linewidth}
    \centering
    \centerline{\includegraphics[width=4.2cm]{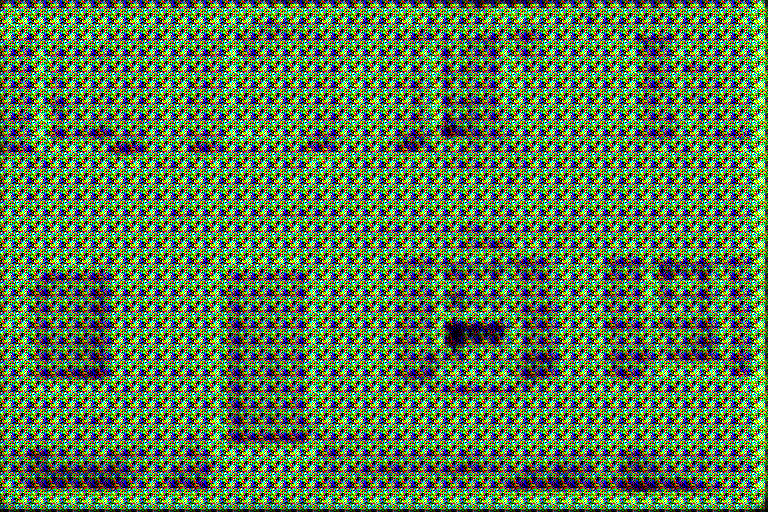}}
    \label{fig_adv_ex_rec}
    \centerline{(b) Adversarial example.}\medskip
  \end{minipage}
  \caption{An ``adversarial" example (\textit{kodim01} image) produced by the compression network. The true VMAF score calculated from (a) and (b) is $5.35$ (which indicates a very poor-quality image), while the proxy network predicts a quality score of $97.74$.}
  \label{fig_adv_example}
  \end{figure}

\section{Related work}
Most of the work on deep image transformation problems has focused on investigating novel network architectures or improving convergence speed. The selection of an appropriate loss function that is consistent with human perception, however, has not been studied much. Here, we review related studies that are closely related to perceptual optimization. As tractable tools, SSIM and MS-SSIM have been widely adopted because of the simple analytical form of their gradients and their computational ease. Moreover, their convexity properties \cite{channappayya08, Brunet2012} makes them feasible targets for optimization. Two recent studies adopted structural similarity functions as loss layers of image generation models, obtaining improved results, as validated by conducting a human subjective study \cite{Snell2017} and by objective evaluation against several other perceptual models \cite{Zhao2017}.

Rather than optimizing a mathematical function, another approach uses a deep neural network to guide the training. Recent experimental studies suggest that the features extracted from a well-trained image classification network have the capability to capture information useful for other perceptual tasks \cite{zhang2018perceptual}. Mathematically, the perceptual loss is defined as
\begin{equation}
\begin{split}
  \mathcal{L}_{\mathrm{perceptual}} 
             =\sum_i \frac{1}{N_i} \abso{ \phi_i\left(\mathbf{x}\right)-\phi_i\left(\mathbf{\hat{x}}\right) }^2,
\end{split}
\end{equation}
where $\phi_i$ denotes the output feature map of the $i$-th layer with $N_i$ elements of a pre-trained network $\phi$. In practice, the loss computed from the high-level features extracted from a pre-trained VGG classification network \cite{SimonyanZ14a}, also called VGG loss, has been commonly adopted for diverse computer vision tasks. The VGG loss has been applied to such diverse tasks as style transfer \cite{JohnsonAF16,GatysEBHS17}, superresolution \cite{BrunaSL15,JohnsonAF16,LedigTHCCAATTWS17,Sajjadi2017}, and image inpainting \cite{Yang_2017_CVPR}. Despite its ubiquity, this ``unreasonable'' perceptual loss is notorious for creating unpleasant artifacts \cite{JohnsonAF16}. Most importantly, this edge-sharpening loss function is incapable of optimizing a network toward a specific quality model.

\begin{figure*}[!ht]
  \centering
  \includegraphics[height=2.65in]{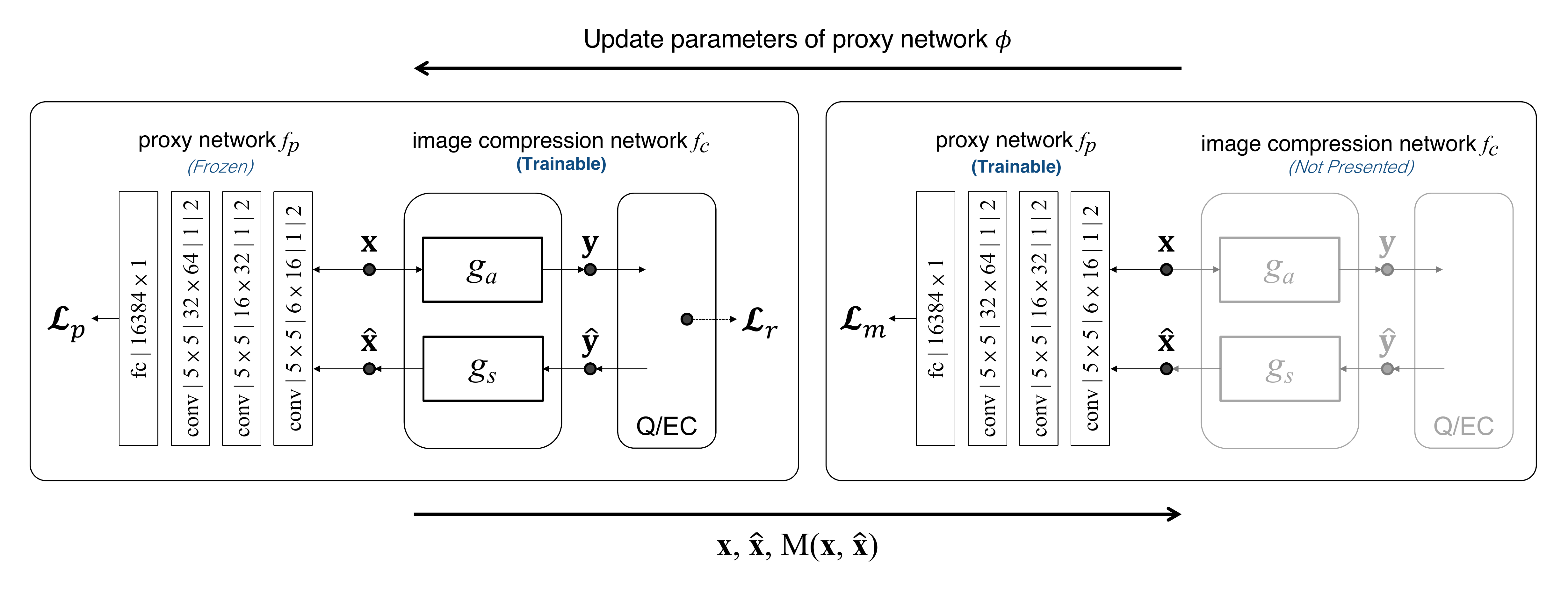}
  \caption{Illustration of the proposed optimization framework. Perceptually training a deep image compression model involves alternating optimization of the compression network $f_c$ and the proxy network of an IQA model $f_p$. Thin arrows indicate the flow of data in the network, while bold arrows represent the information being delivered to update the complementary network. The convolutional parameters of $f_p$ are denoted by ``height $\times$ width $\mid$ input channel $\times$ output channel $\mid$ stride $\mid$ padding''.}
  \Description{The 1907 Franklin Model D roadster.}
  \label{fig_prox_bls}
\end{figure*}

\section{Proposed Perceptual optimization Framework}
\label{sec:ppom}
Learning a successful CNN model depends highly on the size of the training set. Luckily, training a proxy network on an existing model does not require human-labeled subjective quality scores such as mean opinion scores (MOS), which is often the greatest obstacle to learning DNN-based IQA models \cite{Ghadiyaram2016, Kim2017, zqy2020}. Ground truth scores for training the proxy network are easily obtained, given the availability of pristine and distorted patches. To start with, we created a simple network trained by existing datasets \cite{ma2017waterloo} comprising numerous reference and distorted images. Also, the corresponding metric scores were calculated as the ground-truth for training. The proxy network is first learned to predict the metric score given a pristine patch and a distorted patch. Next, the trained proxy network is inserted into the loss layer of the deep compression network with the goal of maximizing the proxy score. Unfortunately, severe complication can arise when applying this straightforward methodology.

We discovered that the deep compression network often generates ``adversarial" examples when its loss layer is the output of a pre-trained network having fixed parameters. Figure~\ref{fig_adv_example} shows such an ``adversarial" example generated by the deep compression network using a proxy network as its loss function. In this example, the proxy network was trained to mimic the VMAF algorithm. However, comparing Fig.~\ref{fig_adv_example}(a) with Fig.~\ref{fig_adv_example}(b), it is apparent that the true VMAF score and the proxy VMAF score predicted by the network $f_p$ are very different. This can be understood by considering the training of the network to be an interpolation problem, whereby the neural networks maps a test image to an accurate quality score. However, when the input is too different from the training set, the proxy network may produce a poor interpolation result. Additionally, as pointed out in \cite{ChengPerception2019}, the conventional distortion types in public domain databases are generally quite different from distortions created by a deep neural networks. In this regard, training a proxy network on previously created databases might be suboptimal for this problem.

\subsection{Alternating Learning Framework}
In order to tackle the aspect, our approach to training an image compression model in a perceptually optimized way is depicted in Fig. \ref{fig_prox_bls}. The idea is to simply use the adversarial examples along with their objective quality scores as additional training data of the proxy network. The proxy network is then updated, enabling it to predict proxy quality much more accurately. This framework involves optimizing an image compression network $f_c$, and a proxy network of an IQA model $f_p$. In each training iteration, the two networks are alternately updated as follows:

\textbf{Deep Compression Network.} To integrate the proxy network $f_p$ into the update of $f_c$ given a mini-batch $\mathbf{x}$, the model parameters of $f_p$ are fixed during training. In order to minimize perceptual distortion, the output of $f_p$ becomes part of the objective in the optimization of $f_c$:
\begin{equation}
  f_p\left( \mathbf{x},\hat{\mathbf{x}} \right) = f_p\left( \mathbf{x},f_c\left( \mathbf{x} \right)\right).
\end{equation}
By back-propagating through the forward model, the loss derivative is used to drive $f_c$.

\textbf{Proxy IQA Network.} Given a mini-batch pair $\mathbf{x}$ and $\hat{\mathbf{x}}$ collected from the most recent update of the compression network, the quality scores $\mathbf{M}(\mathbf{x}, \hat{\mathbf{x}})$ are calculated. The network $f_p$ is updated to optimally fit $\mathbf{M}$ given the input $\left\{\mathbf{x},\hat{\mathbf{x}}\right\}$. Note that the compression network $f_c$ is not needed in this part of the training. As may be seen, $f_p$ is incorporated into the training of the compression network. However, it is important to understand that it is not present during the inference phase.

By applying the proposed alternating training, the proxy network is capable of spontaneously adapting to newly generated adversarial patches. In addition, exotic artifacts created by deep image compression can be ``seen'' by the proxy IQA network: the patches reconstructed by the compression network are directly used to update the proxy network, hence the aforementioned problem becomes immediately resolved.

\begin{table*}[!t]
  \caption{Comparison of conventional codecs and optimized deep image compression: average change of BD-rate expressed as percentage, using three different IQA models to train the compression network. The baseline of comparison is the MSE-optimized BLS model \cite{BalleLS16a}. Smaller or negative values indicate better coding efficiency.}
  \label{tab:comparison}
  \begin{tabular}{l rrrr c rrrr c rrrr }
    \toprule
    Image Dataset  && \multicolumn{2}{c}{Kodak} &&
    && \multicolumn{2}{c}{Tecnick} &&
    && \multicolumn{2}{c}{NFLX Billboard} &\\
    \cmidrule(r){2-5} \cmidrule(r){7-10} \cmidrule(r){12-15}
    BD-rate Metric &PSNR & SSIM & MSIM & VMAF &
                   &PSNR & SSIM & MSIM & VMAF &
                   &PSNR & SSIM & MSIM & VMAF \\
    \midrule
    JPEG                  & 113.99 & 129.49 & 149.86 & 78.36 &
                          & 119.33 & 218.04 & 171.59 & 89.73 &
                          & 102.28 & 143.99 & 168.20 & 89.95 \\
    JPEG2000              & -11.51 & 6.25  & -1.02 & -33.39 &
                          & -13.06 & -1.55 & -8.41 & -34.25 &
                          & -27.81 & 1.43  & -3.93 & -38.98 \\
    HEVC$\mathrm{_{444}}$ & -26.35 & -6.32 & -6.12 & -28.23 &
                          & -28.32 & -8.97 & -11.07& -27.65 &
                          & -49.43 & -17.12& -16.06& -35.03 \\
    HEVC$\mathrm{_{420}}$ & -27.33 & -25.98 & -24.97 & -42.18 &
                          & -19.48  & -28.97 & -33.95 & -46.67 &
                          & -37.63 & -35.41 & -33.88 & -50.91 \\
    \textbf{BLS \SSIMp}   & 15.89  &-21.31 &-19.25 & 7.19  &
                          &  8.67  &-10.79 &-16.11 & 8.68  &
                          & 16.79  &-19.01 &-17.73 & 9.75 \\
     \textbf{BLS \MSIMp}  & 11.67  & -11.58 & -21.77 & -0.17 &
                          & 4.47   & -17.40 & -23.50 & 0.19  &
                          & 12.28  & -11.59 & -23.53 & 4.34 \\
    \textbf{BLS \VMAFp}   & 5.23   & -6.53 & -7.78 & -23.35 &
                          & 6.23   & -8.45 & -5.97 & -23.78 &
                          & 7.00   & -4.35 & -5.43 & -21.97 \\
    BMSHJ MSE \cite{balle2018variational}
                          & -21.46 & -10.94 & -10.17 & -25.78 &
                          & -26.03 & -20.22 & -16.71 & -33.75 &
                          & -36.64 & -21.21 & -21.08 & -38.01 \\
    \textbf{BMSHJ \VMAFp} & -15.90 & -13.57 & -13.17 & -47.11 &
                          & -19.64 & -23.14 & -16.73 & -53.18 &
                          & -29.96 & -18.87 & -19.29 & -56.06 \\
    \bottomrule
  \end{tabular}
\end{table*}

\subsection{Network Architecture}
The proxy IQA network $f_p$ takes a reference patch $x$ and a distorted patch $\hat{x}$ as input, where both have $W\times H$ pixels. They are then concatenated into a 6-channel signal, where a $W \times H \times 6$ raw input $\{x,\hat{x}\}$ is fed into the network and reduced to a predicted quality score. As depicted in Fig.~\ref{fig_prox_bls}, the network $f_p$ may be as simple as a shallow CNN consisting of three stages of convolution, ReLU nonlinearity, and subsampling. The spatial size is reduced by a factor of $2$ after each stage via $2\times2$ max pooling layers. The size of convolution kernels are fixed to $5\times5$ for all stages, while the number of filters at the first stage is set to
$16$ and is increased by a factor of 2 for each subsequent stage. Finally, $64$ feature maps with size $\frac{W}{8}\times \frac{H}{8}$ are flattened and fed to a fully connected layer which yields the output. The parameterization of each layer is detailed in the figure. 

The image compression network comprises an analysis transform ($g_a$) at the encoder side, and a synthesis transform ($g_s$) at the decoder side. Both transform units are implemented as consecutive layers of convolution-down(up) sampling-activation.Instead of the commonly used ReLU, a generalized divisive normalization (GDN) transform is adopted as the activation function \cite{pcs_Balle18}. It is similar to the local gain control behavior in human visual system, where visual signals are normalized by their rectified neighbors. Lastly, the functional block ``Q/EC" denotes quantization and entropy coding. In this work, we build on two different deep image compression models \cite{BalleLS16a,balle2018variational}.

\subsection{Loss Functions}
As illustrated in Fig.~\ref{fig_prox_bls}, let $\mathbf{x}$, $\mathbf{y}$, $\hat{\mathbf{y}}$, and $\hat{\mathbf{x}}$ be the source batch, latent presentation, quantized latent presentation, and reconstructed batch, respectively. The model parameters in the analysis and synthesis transforms are collectively denoted by $\theta=(\theta_a,\theta_s)$. The proxy network $f_p$ has model parameters $\phi$. Given a perceptual metric $M$, the goal is to optimize the full set of parameters $\theta$, $\phi$, such that the learned image codec can generate a reconstructed image $\hat{\mathbf{x}}$ that has a high perceptual quality score $M(\mathbf{x}, \hat{\mathbf{x}})$. Furthermore, the rate should be as small as possible.

Generally, learned image compression network is optimized by minimizing the objective function defined by
\begin{equation}\label{eq:baseline_loss}
  \mathcal{L}\left(\theta\right)=\lambda \mathcal{L}_{d} +\mathcal{L}_{r},
\end{equation}
which has a similar notion as rate-distortion optimization (RDO) in conventional codecs. Under this scheme, $\mathcal{L}_{d}$ is the residual between the source patch and the reconstructed patch mapped by $d(.)$
\begin{equation}\label{eq:pixel_loss}
  \mathcal{L}_{d}\left(\theta\right) =d\left( \mathbf{x}-\hat{\mathbf{x}} \right),
\end{equation}
where $d(.)$ is a distance function such as mean square error or mean absolute error. On the other hand, $\mathcal{L}_{r}$ is the rate loss representing the bit consumption of an encode $\hat{\mathbf{y}}$. We followed the original work in \cite{BalleLS16a}, where the rate loss is defined by
\begin{equation}\label{eq:rate_loss}
  \mathcal{L}_{r}\left(\theta \right)
  = -\log_2p_{\hat{\mathbf{y}}}\left( \hat{\mathbf{y}} \right).
\end{equation}
The term $p_{\hat{\mathbf{y}}}\left( \hat{\mathbf{y}} \right)$ denotes the entropy model. This entropy term is minimized when the actual marginal distribution and $\hat{\mathbf{y}}$ have the same distribution. During training, the latent presentation $\mathbf{y}$ is quantized to $\hat{\mathbf{y}}$ by adding i.i.d uniform noise $\mathcal{U}\left(-\frac{1}{2},\frac{1}{2}\right)$. Then, $\hat{\mathbf{y}}$ is used to estimate the rate via (\ref{eq:rate_loss}). Unlike the training phase, normal rounding-based quantization is applied to quantize $\mathbf{y}$. Then, entropy coders such as variable length coding or arithmetic coding can be used to losslessly encode the discrete-valued data into the bitstream during the inference.

Rather than just minimizing an $\ell_p$ norm between $\mathbf{x}$ and $\hat{\mathbf{x}}$, we introduce a loss term $\mathcal{L}_{p}$. This proxy loss $\mathcal{L}_{p}$ is defined to maximize the output of proxy network $f_p$, denoted by $\hat{M}$, with fixed network parameters $\phi$:
\begin{equation}\label{eq:proxy_loss}
  \mathcal{L}_{p}\left(\theta ; \phi\right)
  =M_{max} - \hat{M}\left( \mathbf{x},\hat{\mathbf{x}} \right).
\end{equation}
Here $M_{max}$ denotes the upper bound of the model $M$, which is a constant to the loss function. Finally, the total loss function for optimizing the compression network is the weighted combination of the losses:
\begin{equation}\label{eq:total_loss}
  \mathcal{L}_{t}\left(\theta ; \phi\right)
  =\lambda\left[ \alpha\mathcal{L}_{p} + \left( 1-\alpha \right) \mathcal{L}_{d} \right] + \mathcal{L}_{r},
\end{equation}
where $\lambda$ controls the trade-off between bitrate and distortion of the encoded bitstream, and $\alpha$ weights $\mathcal{L}_{p}$ against $\mathcal{L}_{d}$. Here, the term $\mathcal{L}_{d}$ plays a different role as a regularization term. Since the proxy network is updated at each step, the loss function $\mathcal{L}_{p}$ is also changed. The pixel loss serves to stabilize the training process. In our model, we empirically set $\alpha=1.54e-3$ and $M_{max}=100$ when optimizing for VMAF.

The proxy network $f_p$ aims to mimic an image quality model $M$. While updating $f_p$, we define a metric loss $\mathcal{L}_{m}$ to attain this objective given two image batches $\mathbf{x}$ and $\mathbf{\hat{x}}$:
\begin{equation}\label{eq:metric_loss}
  \mathcal{L}_{m}\left(\phi;\hat{\mathbf{x}}\right)
  = \abso{ \hat{M}\left( \mathbf{x},\hat{\mathbf{x}} \right) - 
  M\left( \mathbf{x},\hat{\mathbf{x}} \right) }^2.
\end{equation}
Note that $\hat{\mathbf{x}}$ is a constant, since it is obtained from the reconstructed patches generated during the most recent update of the compression network.

\begin{figure}[!t]
  \centering
  \includegraphics[height=2.4in]{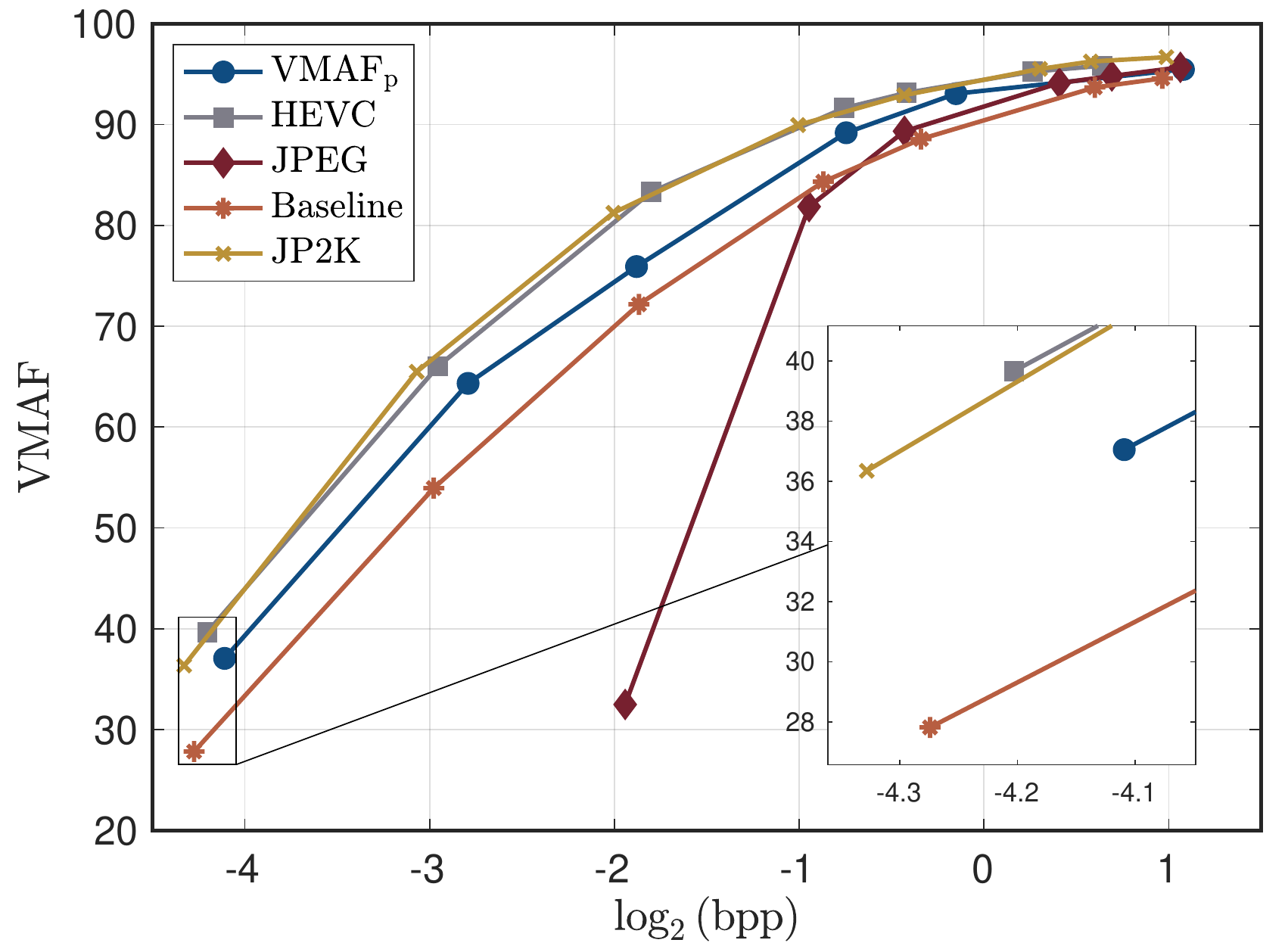}
  \caption{Rate-Distortion (RD) curves for different image compression algorithms on the image \textit{kodim10}, measured with VMAF.}
  \Description{The 1907 Franklin Model D roadster.}
  \label{fig_rdkodim10}
\end{figure}

\begin{figure}[!t]
	\centering
	\renewcommand{\tabcolsep}{1.6pt} 
	\def\imgwid{0.27\textwidth}
	\def\rdhei{0.223\textwidth}
	\def\imghei{0.199\textwidth}
  \def\imgheid2{0.088\textwidth}
	\def\rdheid2{0.18\textwidth}
	\def\rd_shift{-0.123\textwidth}
  \def\im_shift{-0.102\textwidth}

	\begin{tabular}{ccccc}
        Source &
        BLS-MSE &
        JPEG2000 &
        BLS-\VMAFp &
        HEVC$\mathrm{_{444}}$ \\
        \includegraphics[height=\imgheid2]{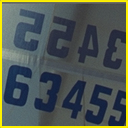} & 
        \includegraphics[height=\imgheid2]{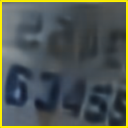} & 
        \includegraphics[height=\imgheid2]{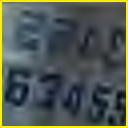} &
        \includegraphics[height=\imgheid2]{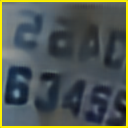} & 
        \includegraphics[height=\imgheid2]{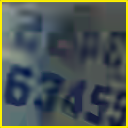} \\
		    \includegraphics[height=\imgheid2]{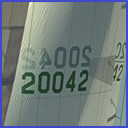} &
        \includegraphics[height=\imgheid2]{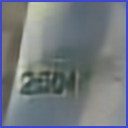} &
        \includegraphics[height=\imgheid2]{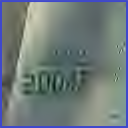} &
        \includegraphics[height=\imgheid2]{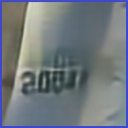} &
        \includegraphics[height=\imgheid2]{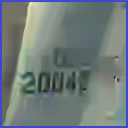} \\
                         bpp\,/\,VMAF 
                         & 0.052\,/\,27.82 
                         & 0.050\,/\,36.39 
                         & 0.058\,/\,37.06 
                         & 0.054\,/\,39.67 \\
                         PSNR\,/\,SSIM 
                         & 26.12\,/\,0.817 
                         & 26.53\,/\,0.806 
                         & 26.32\,/\,0.840 
                         & 27.78\,/\,0.843 \\
	\end{tabular}
	\caption{Visualization of decoded pictures (\textit{kodim10}) from differend compression models. Each cropped to $128\times 128$ patch for display purpose.}
	\label{fig:compare_lin2}
\end{figure}

\section{Experiments}
\label{sec:experiments}
The following subsections thoroughly describe the experimental setup. We also present the quantitative evaluation and subjective comparison.
\subsection{Experimental Setup}
\textbf{Implementation Details.} We used the \texttt{TensorFlow} framework (version 1.12) to implement the proposed method. The Adam solver \cite{kingma:adam} were used to optimize both the proxy network and the deep compression network, with parameters $(\beta_1, \beta_2)=(0.9,0.999)$ and a weight decay of $0.01$. We set the initial learning rates for both networks at fixed values of $1e-4$ for the first 2M steps and a lower learning rate of $1e-5$ for an additional 100K steps. Thus, the networks were trained on 2.1M iterations of back-propagation. To fairly compare deep image compression models having different loss layers, we used $192$ filters at every layer, and trained all of the models using the same number of steps. All of the models were trained using NVIDIA 1080-TI GPU cards.

\textbf{Training Setup.} We used a subset of the $6507$ processed images from the ImageNet database \cite{imagenet_cvpr09} as training data. As described in \cite{BalleLS16a}, small amounts of uniform noise were added to the images. The images were then down-sampled by random factors to reduce compression artifacts and high-frequency noise, and randomly cropped to a size of $256\times256$. In each mini-batch, we randomly sampled $8$ image patches from the subset. We then cropped the images to $128\times128$ patches resulting in an $8\times 128\times 128\times 3$ tensor.

\textbf{Evaluation Datasets.} To evaluate various image codecs, we utilized the Kodak dataset of $24$ very high quality uncompressed $768\times512$ images. This publicly available image set is commonly used to evaluate image compression algorithms and IQA models. We also used a subset of the Tecnick dataset \cite{stag_tecnick} containing $100$ images of resolution $1200\times1200$, and $223$ billboard images collected from the Netflix library \cite{Sinno2020}, yielding images having more diverse resolutions and contents. It should be noted that none of these test images were
included in the training sets, to avoid overfitting problems.

\textbf{Evaluation Setup.} As is the common practice in the field of video coding, we measured the objective coding efficiency of each image codec using the Bj\o{}ntegaard-Delta bitrate (BD-rate) \cite{BDRate01}, which quantifies average differences in bitrate at the same distortion level relative to another reference encoder. To calculate BD-rate, we encoded the images at eight different bitrates, ranging from $0.05$ bpp (bit per pixel) to $2$ bpp. In all the experiments conducted, we denote the image compression model \cite{BalleLS16a} as the BLS model. The performances of all of the codecs were compared to the same baseline -- the MSE-optimized BLS model. A negative number of BD-rate means the bitrate was reduced as compared with the baseline. The input image formats used were YUV444 for JPEG and JPEG2000, and both YUV420/444 for intra-coded HEVC, respectively. Lastly, the distortion levels that were used for BD-rate calculation were quantified using PSNR, SSIM, MS-SSIM (also represented by MSIM in the table), and VMAF.

\subsection{Comparison with Different Codecs}
We comprehensively evaluated perceptual deep compression using different perceptual optimization protocols (highlighted in boldface), against three conventional image codecs: JPEG, JPEG2000, and intra coding of HEVC. Extensive experiments were carried out using three perceptual IQA models as optimization targets. Table~\ref{tab:comparison} tabulates the benchmark study on the aformentioned three datasets. Each cell shows the BD-rate relative to the BLS baseline, with respect to different quality models. We denote an optimized compression model for a given IQA model $M$ using (\ref{eq:total_loss}) and (\ref{eq:metric_loss}) by $M\mathrm{_p}$. In addition to the BLS model, we also deployed the proposed \VMAFp~optimization framework on a more sophisticated deep compression model \cite{balle2018variational} (BMSHJ) to test its generality. We report the BD-rate changes obtained, averaged over all the images in each dataset. These results show that our optimization approach is able to successfully optimize a deep image compression model over different IQA algorithms. Indeed, significant BD-rate reductions were obtained in many cases. An interesting observation can be made that, unlike using other IQA models used as targets of the proposed optimization, \VMAFp~optimization delivers coding gain with respect to all of the BD-rate measurements, except the PSNR BD-rate. This suggests VMAF being a good optimization target. 

As a basic test, we subjectively compare results yielding similar bitrates but different objective quality scores. Figure \ref{fig:compare_lin2} shows a visual comparison under extreme compression (around 0.05 bpp). Obviously, the \VMAFp-optimized model significantly outperformed the MSE-optimized baseline model, delivering performance comparable to HEVC and JPEG2000 with respect to VMAF score and subjective quality. We also plot the corresponding VMAF Rate-distortion (RD) curve, a common tool for comparing different encoders, in Fig. \ref{fig_rdkodim10}. We observe that the proposed optimization scheme generally leads to a compression gain in VMAF. At high bitrates, the \VMAFp-optimized model yielded comparable VMAF scores as the baseline MSE-optimized model, while consuming significant fewer bits. In this particular example, roughly $16\%$ of the bits can be reduced without suffering perceptual quality.

\subsection{Study of Alternating Training}
\begin{figure}[tb]
  \begin{minipage}[b]{0.9\linewidth}
  \centering
  \centerline{\includegraphics[height=1.36in]{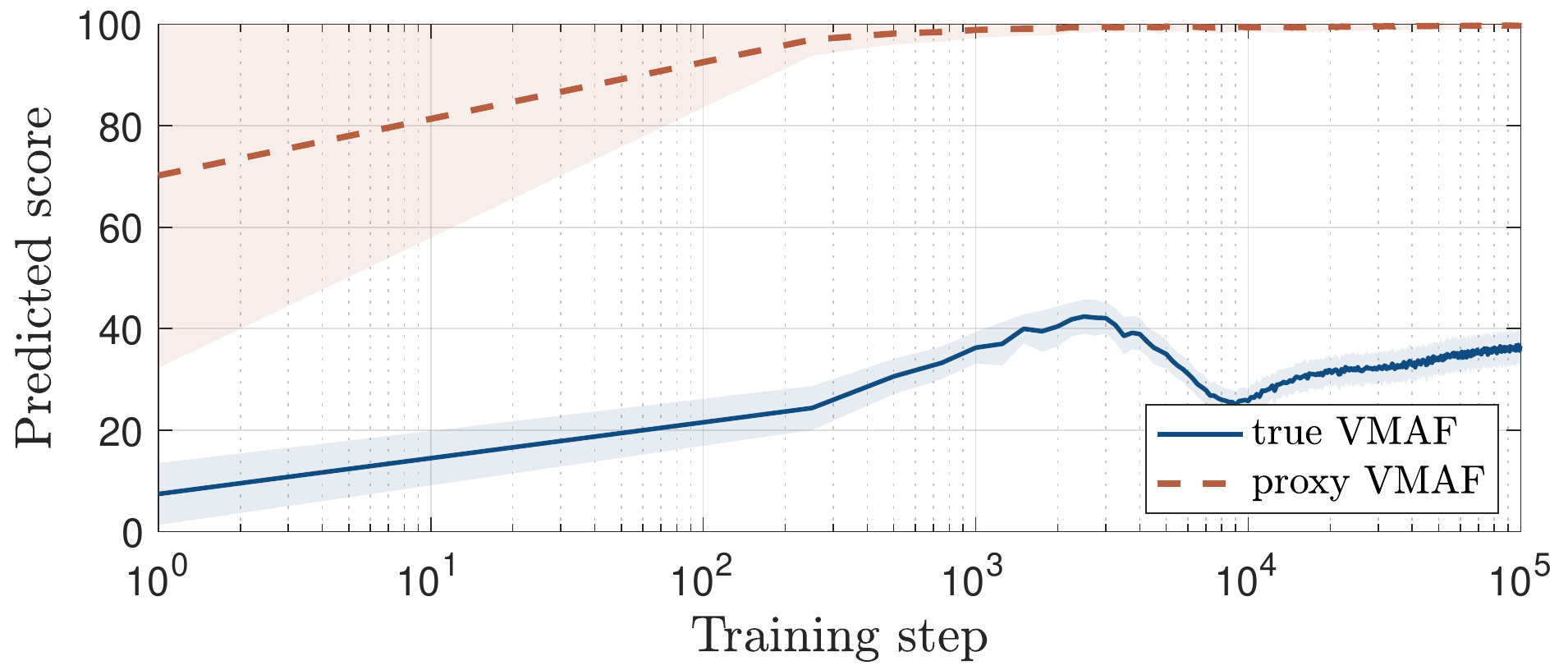}}
  \label{fig_train_pre}
  \centerline{(a) Model learned with a pre-trained proxy network.}\medskip
  \end{minipage}
  \hfill
  \begin{minipage}[b]{0.9\linewidth}
  \centering
  \centerline{\includegraphics[height=1.36in]{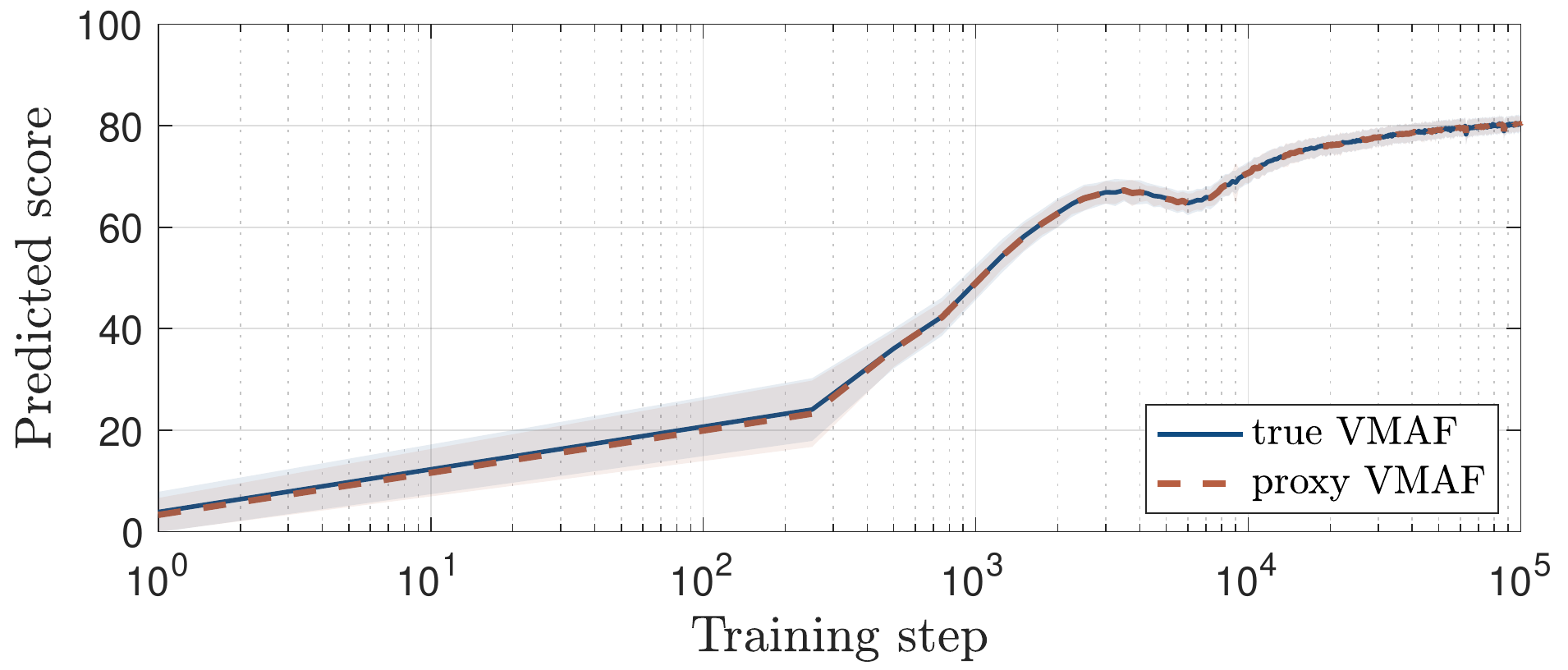}}
  \label{fig_train_alt}
  \centerline{(b) Model learned from the proposed alternating training process.}\medskip
  \end{minipage}
  \caption{Comparison of two different optimization strategies during the training process. We plot true VMAF scores and proxy VMAF scores (predicted by the proxy network) of the reconstructed batch. The two scores are plotted in mean values (lines) and one standard deviations (shadows).}
  \label{fig_train_plt}
\end{figure}

To further understand the behavior of the prpopsed alternating training, we compared true VMAF scores against proxy VMAF scores in Fig. \ref{fig_train_plt}. All of the scores were calculated on the reconstructed patches produced during training. Figure \ref{fig_train_plt}(a) shows that the proxy VMAF scores quickly approached $100$, whereas much lower true VMAF scores were assigned to the patches produced by the compression model. This directly reflects the problem we have mentioned (Sec. \ref{sec:ppom}) when a pre-trained proxy network is applied. However, when the reconstructed patches were feed into the proxy network along with their objective quality scores, the proxy network is updated straightaway to predict proxy quality much more accurately. As shown in Fig. \ref{fig_train_plt}(b), the true and proxy scores become highly consistent early in the training process.

\subsection{Computational Cost}
It is critical for a learned image compression model to have comparable execution time to other codecs. We compiled the source code of standard codecs, in order to be able to compare them on the same computer with a $2.10$GHz CPU and $4$ GTX-1080TI GPUs. The results were then calculated by averaging the runtime over all $24$ Kodak images under different bitrate settings. The encoding and decoding times of the various compared codecs are reported in Table \ref{tab:freq}. It may be observed that the time complexity of the MSE-optimized and \VMAFp-optimized models are nearly identical, as they deploy the same network architecture in application. Overall, the runtime of learned image compression models are acceptable and can be further reduced if performed on a GPU. It is worth noting that the BLS models achieves the fastest decoding speed when a GPU is available. It should also be noted that the decoding time of HEVC was estimated from the reference software \texttt{HM16.9}, which might be slow.

\begin{table}
  \caption{Average processing speed in milliseconds for different compression models. Model loading time for deep compression is excluded.}
	\renewcommand{\tabcolsep}{3.5pt} 
	\renewcommand{\arraystretch}{0.95} 
  \label{tab:freq}
  \begin{tabular}{lcrr}
    \toprule
    Compression model & & Encode time & Decode time\\
    \midrule
    JPEG                                & CPU & 43.02   & 62.88          \\
    JPEG2000                            & CPU & 10.80   & 36.79          \\
    HEVC                                & CPU & 4578.57 & 89.88          \\
    BLS MSE
    & CPU & 251.01  & 117.93         \\
    & \textit{GPU} & \textit{231.62} 
    & \textit{32.56}                 \\
    BLS \VMAFp
    & CPU & 246.57  & 119.02         \\
    & \textit{GPU} & \textit{229.26}
    & \textit{29.22}                 \\
    BMSHJ MSE
    & CPU          & 351.23          & 378.68 \\
    & \textit{GPU} & \textit{312.28} & \textit{344.13} \\
    BMSHJ \VMAFp   
    & CPU          & 367.71          & 380.04 \\
    & \textit{GPU} & \textit{308.53} & \textit{341.96} \\
  \bottomrule
\end{tabular}
\end{table}

\section{Conclusion}
\label{sec:conclusion}

In this work, we focus on designing the loss function for deep image compression. In particular, we have presented a framework for perceptually optimizing a generative network and a proxy image quality assessment network. When integrated into deep image compression models, our method allows end-to-end training and can provide compression gains with respect to different IQA metrics. We believe that the idea behind the proposed training framework is general. With proper modifications of the framework parameters or the architecture of the proxy network, the approach has the potential to improve on a wide variety of image restoration problems with weak MSE based ways of optimization.


\begin{acks}
  This research is supported by Netflix. The authors thank the Texas Advanced Computing Center (TACC) at The University of Texas at Austin for providing HPC resources that have contributed to the research results reported within this paper. URL: \url{http://www.tacc.utexas.edu}
\end{acks}

\bibliographystyle{ACM-Reference-Format}
\bibliography{sample-base}

\end{document}